\documentclass[reprint,
superscriptaddress,
citeautoscript,
amsmath,amssymb,
aps,
prb
]{revtex4-1}

\usepackage[pdftex]{graphicx}
\usepackage{tabularx}
\usepackage{dcolumn}
\usepackage{bm}
\usepackage{hyperref}
\usepackage{color}

\begin{document}

\preprint{APS/123-QED}

\title{Modifications of the Meissner screening profile in YBa$_2$Cu$_3$O$_{7-\delta}$ thin films\\ by gold nanoparticles}

\author{E. Stilp}
 \affiliation{Laboratory for Muon Spin Spectroscopy, Paul Scherrer Institut, CH-5232 Villigen PSI, Switzerland}
 \affiliation{Physik-Institut der Universit\"at Z\"urich, Winterthurerstrasse 190, CH-8057 Z\"urich, Switzerland}
 
\author{A. Suter}
 \affiliation{Laboratory for Muon Spin Spectroscopy, Paul Scherrer Institut, CH-5232 Villigen PSI, Switzerland}

\author{T. Prokscha}
 \affiliation{Laboratory for Muon Spin Spectroscopy, Paul Scherrer Institut, CH-5232 Villigen PSI, Switzerland}

\author{Z. Salman}
 \affiliation{Laboratory for Muon Spin Spectroscopy, Paul Scherrer Institut, CH-5232 Villigen PSI, Switzerland}

\author{E. Morenzoni}
 \affiliation{Laboratory for Muon Spin Spectroscopy, Paul Scherrer Institut, CH-5232 Villigen PSI, Switzerland}

\author{H. Keller}
 \affiliation{Physik-Institut der Universit\"at Z\"urich, Winterthurerstrasse 190, CH-8057 Z\"urich, Switzerland}
 
\author{C. Katzer}
 \affiliation{Institut f\"ur Festk\"orperphysik, Friedrich-Schiller-Universit\"at, Helmholtzweg 5, D-07743 Jena, Germany}
 
\author{F. Schmidl}
 \affiliation{Institut f\"ur Festk\"orperphysik, Friedrich-Schiller-Universit\"at, Helmholtzweg 5, D-07743 Jena, Germany}

\author{M. D\"obeli}
 \affiliation{Laboratory of Ion Beam Physics, ETH-Z\"urich, Schafmattstrasse 20, CH-8093 Z\"urich, Switzerland}

\date{\today}

\begin{abstract}
Adding Au nanoparticles to YBa$_2$Cu$_3$O$_{7-\delta}$ thin films leads to an increase of the superconducting transition temperature $T_{\rm c}$ and the critical current density $j_{\rm c}$. While the higher $j_{\rm c}$ can be understood in terms of a stronger pinning of the flux vortices at the Au nanoparticles, the enhanced $T_{\rm c}$ is still puzzling. In the present study, we determined the microscopic magnetic penetration profiles and the corresponding London penetration depths $\lambda_{\rm L}$ in the Meissner state of optimally doped YBa$_2$Cu$_3$O$_{7-\delta}$ thin films with and without Au nanoparticles by low-energy muon spin rotation. By Rutherford backscattering spectrometry, we show that the Au nanoparitcles are distributed over the whole thickness of the thin-film samples. The superfluid density $n_{\rm s} \propto 1/\lambda_{\rm L}^2$ was found to increase in the films containing Au nanoparticles. We attribute this increase of $n_{\rm s}$ to a reduction of the defect density possibly due to defect condensation at the Au nanoparticles.
\end{abstract}

\maketitle
The influence of disorder on cuprate systems is still an open issue. It is known that disorder directly affects the superconducting properties of these systems, but the underlying mechanisms are not fully clarified~\cite{Alloul09}. In YBa$_2$Cu$_3$O$_{7-\delta}$ (\mbox{YBCO}), for instance, the charge carrier transfer from the Cu-O chains to the CuO$_2$ planes is crucial for the appearance of superconductivity. This charge transfer depends strongly on the oxygen order and on the oxygen mobility within the Cu-O chains~\cite{Aligia94}. Substitution of chain Cu by other metal ions leads generally to a suppression of the superconducting transition temperature $T_{\rm c}$~\cite{Shlyk02}. Surprisingly, this does not apply to Au. Superconducting quantum interference device (SQUID) and neutron diffraction measurements performed by Cieplak \textit{et al.}~\cite{Cieplak1990,Cieplak1990b} show that it is possible to incorporate $10\,at.$\% Au into the structure of polycrystalline YBCO, where Au exclusively replaces the chain Cu(1) atoms. This leads to an increase of $T_{\rm c}$ of about $1.5$\,K and to an increase of the $c$-axis lattice constant. The Cu-O bond length between Cu(1) and the oxygen in the Cu-O chains as well as  between  Cu(1) and the bridging oxygen are enlarged. Previous studies revealed also a lower normal state resistance and a minor influence on the normal state magnetization and on the upper critical field~\cite{Welp93}.

Recently, it has been demonstrated that Au nanoparticles, incorporated in YBCO thin films, lead to an increase of $T_{\rm c}$ and the critical current density $j_{\rm c}$ (from $4 \cdot 10^7$\,A/cm$^2$ to $6 \cdot 10^7$\,A/cm$^2$ at $10$\,K)~\cite{Michalowski12,Katzer11}. Since \mbox{YBCO} is superconducting above the boiling temperature of nitrogen this is of special interest for applications. A stronger flux pinning of the vortices at the Au nanoparticles causes a higher $j_{\rm c}$, which is crucial in order to improve superconducting magnetic field sensors~\cite{Katzer11,Katzer12} as used in dc-SQUID gradiometers.

Until now little is known about the effect of the inclusion of Au nanoparticles on the microscopic superconducting parameters, such as the superfluid density $n_{\rm s}$. In the present study we investigated two sets of YBCO thin films with different Au contents by means of low-energy muon spin rotation (LE-$\mu$SR) experiments and Rutherford backscattering spectrometry (RBS). The amount and the distribution of the Au nanoparticles were determined as well as the microscopic changes of the Meissner screening profile and the London penetration depth $\lambda_{\rm L}$. This is of special interest, since $\lambda_{\rm L}$ is related to the superfluid density $n_{\rm s}$:
\begin{equation}
\lambda_{\rm L}^2 \propto \frac{m^{*}}{ n_{\rm s}}, 
\end{equation}
where $m^{*}$ is the effective mass of the supercarriers.

The optimally doped YBCO thin films studied in this work were prepared by pulsed-laser-deposition (PLD) using a KrF excimer laser ($\lambda=248$\,nm, $\tau=25$\,ns) at the Friedrich-Schiller-Universit\"at in Jena. Single-crystal SrTiO$_3$ \mbox{$10\times10\times1$mm$^3$} in size and polished with the surface perpendicular to the $[001]$ crystal axis was used as substrate. In order to realize clustered Au nanoparticles, an initial gold seed layer of $1.8$\,nm thickness was sputtered on the substrates by dc-magnetron sputtering ($I=10$\,mA, $U=215$\,V, growth rate: $4$\,nm/min, $p_{\rm Ar}=5$\,mbar). Subsequently, the Au coated substrates were heated to $780^{\circ}$C to achieve a dewetting of the Au seed layer. The resulting nanoparticles were overgrown by YBCO in an in situ process. The YBCO deposition was realized by a PLD process in an oxygen atmosphere (repetition rate: $5$\,Hz, laser fluence: $2.2$\,J/cm$^2$, growth rate: $20$\,nm/min, $p_{{\rm O}_2}=0.5$\,mbar). In order to obtain the orthorhombic phase, the resulting YBCO thin films were cooled down to room temperature at a rate of $50$\,K/min in pure oxygen ($p_{{\rm O}_2}=800$\,mbar). The resulting  optimally doped YBCO thin films have a typical thickness of $120-200$\,nm and are $c$-axis oriented. With the Au seed layer thickness used the typical Au particle size is in the range $10-40$\,nm~\cite{Katzer11}. The area distribution of the Au nanoparticles is quite homogeneous, whereas the concentration within the depth profile varies, as later discussed. More details on the growth and the properties of the used samples can be found in Refs.~\onlinecite{Michalowski12}~and~\onlinecite{Katzer11}. The mean distance between the Au nanoparticles is about $50$\,nm. The samples are therefore still in the clean limit, since the correlation length is of the order of $2$\,nm.

\begin{table*}
\caption{Summary of the parameters determined for the investigated optimally doped YBCO thin-film samples with (YBCO$_{\rm Au}$) and without Au nanoparticles (YBCO). The superconducting transition temperatures $T_{\rm c}$ were obtained from resistivity measurements. The average film thicknesses $\overline{d}$ were extracted from RBS measurements. From a global fit of the LE-$\mu$SR data the superconducting thicknesses $d_{\rm sc}$, the dead layer thicknesses to the vacuum interface $z_0$, and the London penetrations depths $\lambda_{ab}$($10$\,K) were determined. $\lambda_{ab}$($0$\,K) and the superconducting gap $\Delta_0$ were extracted from a linear fit in the range $5-35$\,K of $\lambda_{ab}(T)$ measured with LE-$\mu$SR. For comparison single crystal values~\cite{Hossain12} are also presented.}
\label{tab:Tabble1} 
\centering 
\begin{tabularx}{\textwidth}{XXXXXp{2.5cm}XX}
\hline
\hline
Sample & $T_{\rm c}$ (K) & $\overline{d}$ (nm) & $d_{\rm sc}$ (nm) & $z_0$ (nm) & $\lambda_{ab}$($10$\,K) (nm) & $\lambda_{ab}$($0$\,K) (nm) & $\Delta_0$ (meV) \\ \hline
{YBCO film} & $90.2(2)$ & $150(8)$ & $119(1)$ & $0(2)$ & $161(1)$ & $158(1)$ & $22(2)$ \\
{YBCO$_{\rm Au}$ film} & $90.8(1)$ & $152(8)$ & $133(1)$ & $1(2)$ & $154(1)$ & $150(1)$ & $24(2)$ \\
YBCO bulk~\cite{Hossain12} & $94.1(1)$ & & & $10.3(5)$ & & $115(4)$ & $20(4)$ \\
\hline
\hline
\end{tabularx}
\end{table*}

\begin{figure}[b]
	\centering
		\includegraphics[width=\columnwidth]{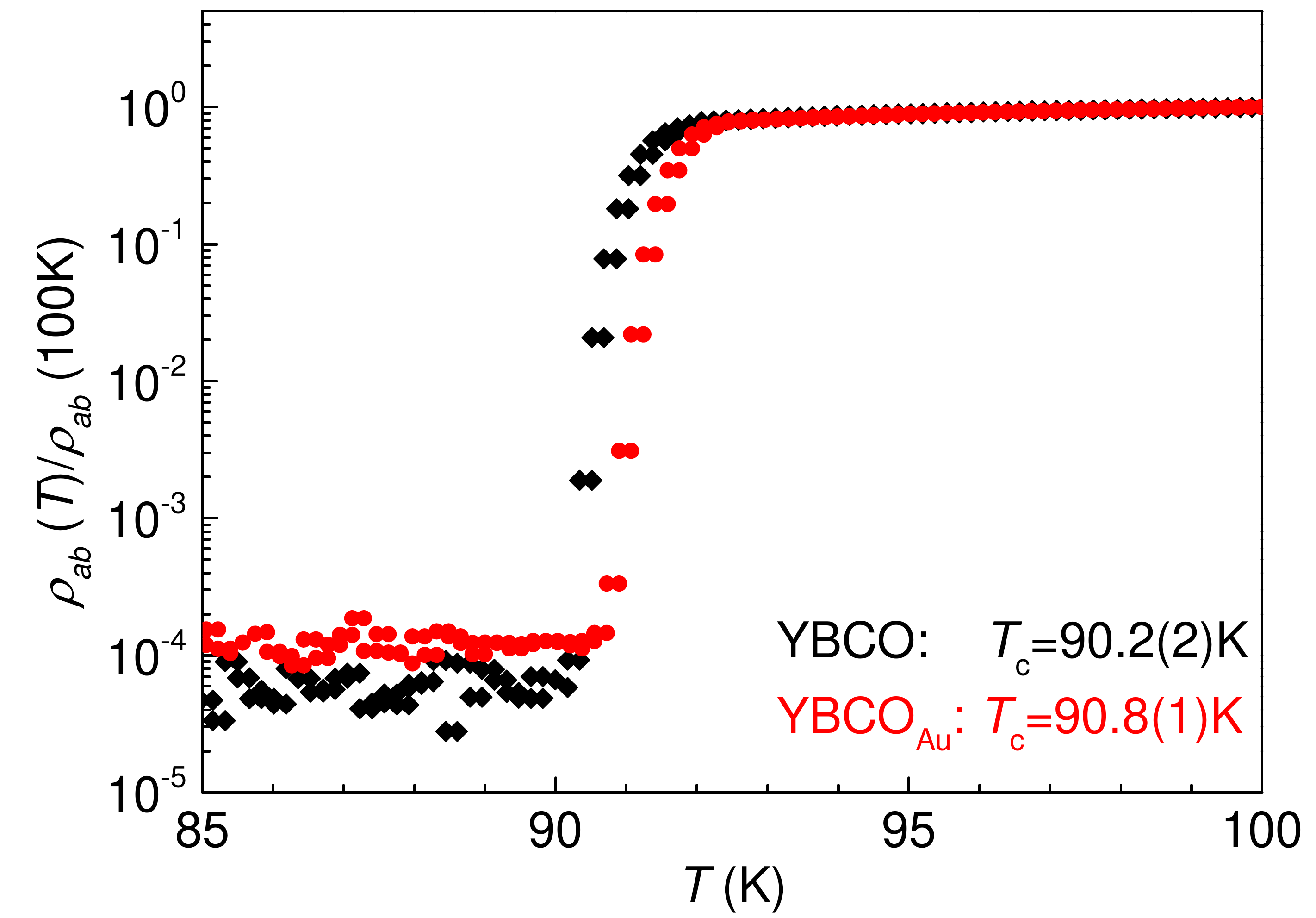}
	\caption{(Color online) In-plane resistivity $\rho_{ab}$ normalized at $100$\,K versus temperature $T$ of PLD-grown optimally doped YBCO thin-film samples: \mbox{YBCO} (black diamonds) and \mbox{YBCO$_{\rm Au}$} (red circles). The listed values of the superconducting transition temperatures $T_{\rm c}$ are defined by $\rho_{ab}(T_{\rm c})/\rho_{ab}(100\,{\rm K}) \leq 2.5 \cdot 10^{-4}$. They are averaged over five measurement cycles (heating and cooling).}
	\label{fig:RversusT}
\end{figure}

Here, we studied two sample sets with (\mbox{YBCO$_{\rm Au}$}) and without (\mbox{YBCO}) Au nanoparticles. As expected, slightly higher $T_{\rm c}$ values were observed by resistivity measurements for the YBCO$_{\rm Au}$ compared to the pristine YBCO thin-film samples (see Fig.~\ref{fig:RversusT}). In order to monitor precisely the amount as well as the distribution of Au as a function of depth in the thin-film samples, RBS measurements were performed at ETH Z\"urich. For RBS a $^4$He beam with an energy of $2-5$\,MeV and a silicon PIN diode detector at a scattering angle of $168^{\circ}$ were used~\cite{Doebeli98}. The elemental composition was determined by using the RUMP software~\cite{Doolittle86}.
\begin{figure}[b]
	\centering
	    \includegraphics[width=\columnwidth]{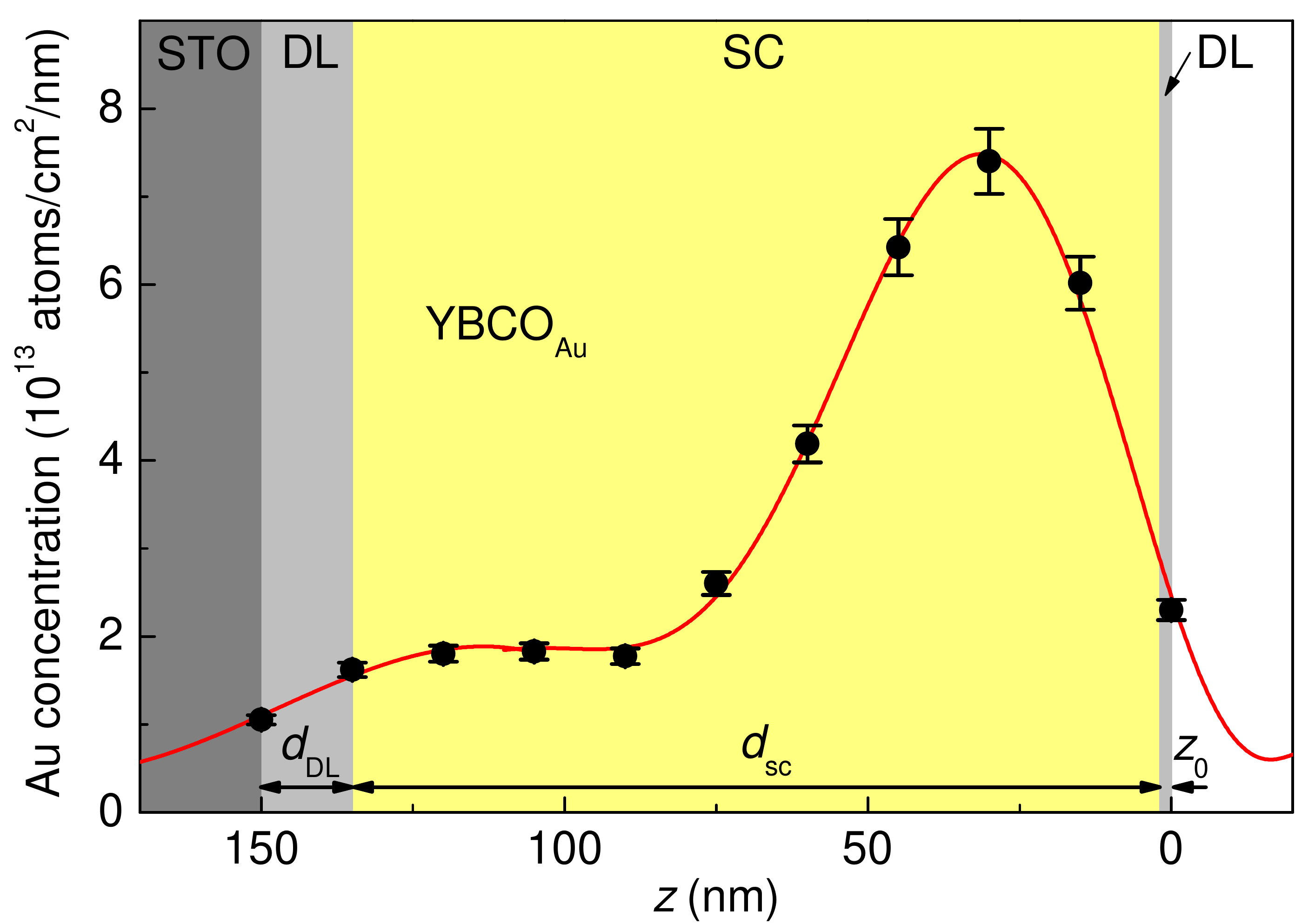}  		
	\caption{(Color online) Gold concentration versus depth $z$ along the crystallographic $c$-axis for a \mbox{YBCO$_{\rm Au}$} thin-film sample determined by RBS. The different regions are labeled: STO is the substrate region (dark gray), DL is the dead layer at the substrate ($d_{\rm DL}$) and at the vacuum ($z_0$) interface (light gray), SC is the superconducting region with thickness $d_{\rm sc}$ (yellow). The red line is a guide to the eye.}
	\label{fig:RBS}
\end{figure}
In the \mbox{YBCO} samples no Au peak is present in the RBS spectra as expected. The stoichiometry of optimally doped YBCO was confirmed within experimental uncertainty. The Au concentrations as function of depth were determined in the \mbox{YBCO$_{\rm Au}$} samples as depicted in Fig.~\ref{fig:RBS}. In these samples the Au nanoparticles are distributed over the whole superconducting region. On average there are about $4(1)\cdot10^{15}$ Au atoms/cm$^2$ in this region. The so-called dead layer close to the substrate interface $d_{\rm DL}$ contains less nanoparticles and exhibits no macroscopic Meissner screening. This is consistent with the lower $T_{\rm c}$ generally observed in this region in these thin-film samples~\cite{Schneidewind95}.

\begin{figure}[b]
	\centering
		\includegraphics[width=\columnwidth]{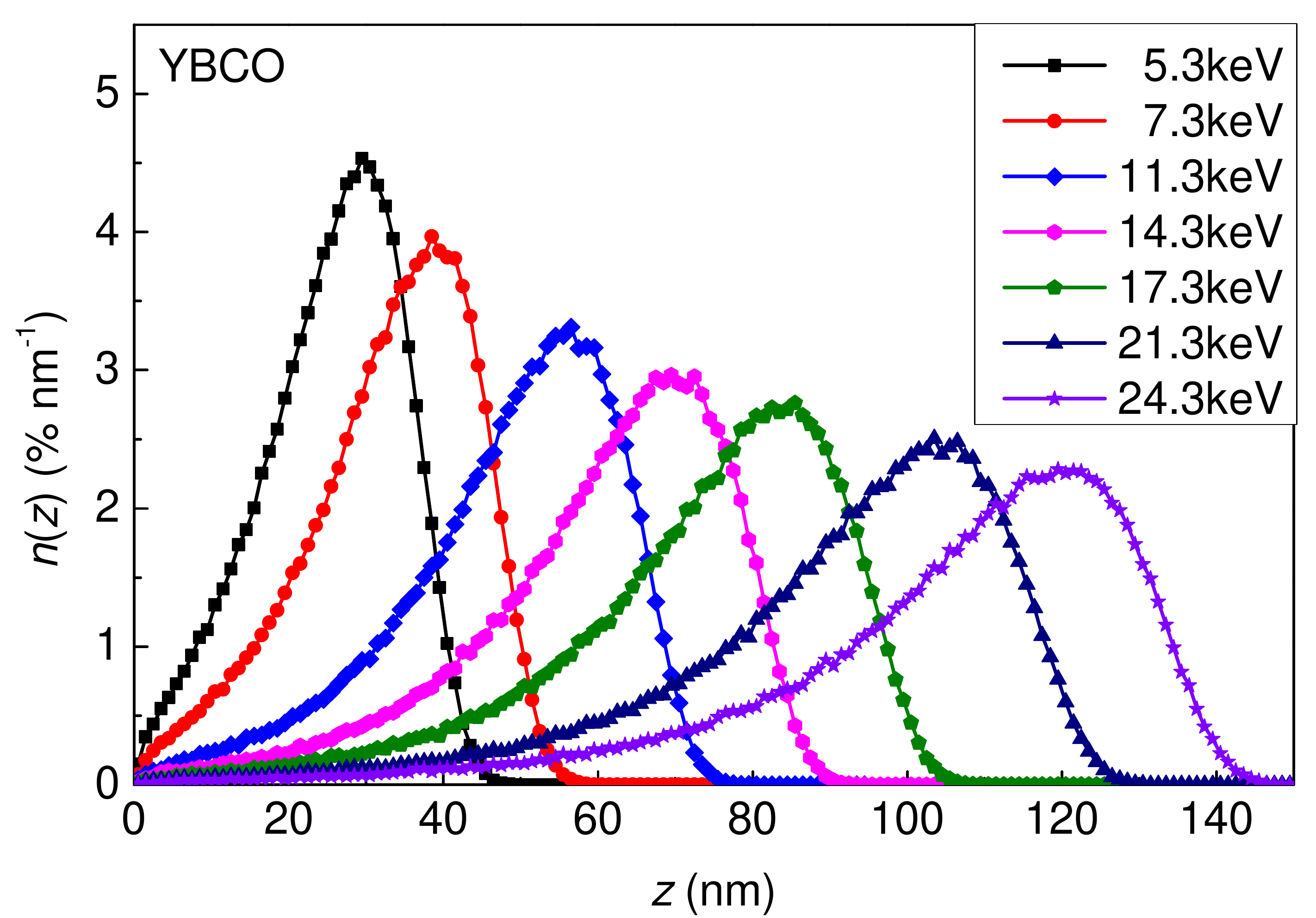}
	\caption{(Color online) The normalized stopping distribution of positively charged muons ${n(z)}$ for different implantation energies (see inset) of a $150$\,nm thick YBCO film simulated with TRIM.SP~\cite{Eckstein91}. The lines are guides to the eye.}
	\label{fig:Profile}
\end{figure}

In order to determine the Meissner screening profile~$B(z)$ and the temperature dependence of the magnetic penetration depth~$\lambda_{\rm L}$ in a straight forward way LE-$\mu$SR experiments were performed at the $\mu$E4 beamline at the Paul Scherrer Institut (PSI, Switzerland)~\cite{Morenzoni04,Prokscha08}. A mosaic of four samples from one set was glued with silver paint on a nickel coated aluminum plate. The low-energy muons ($\mu^+$) are produced at a rate of about $10^4$\,s$^{-1}$ using a solid Ar/N$_2$ moderator. After reacceleration the final energy of the muons is adjusted by a bias voltage at the sample. By tuning the energy between $5$ and $25$\,keV the mean stopping depth of the muons can be varied in a range of $25-106$\,nm in YBCO (see Fig.~\ref{fig:Profile}).

In condensed matter, muons act as local magnetic probes. Measuring the time difference $t=t_{\rm e} - t_{\rm s}$ between the implantation time $t_{\rm s}$ and the decay time $t_{\rm e}$ of the $\mu^+$, detected via the decay positron, allows one to determine the temporal evolution of the muon-spin polarization $P(t)$~\cite{Yaouanc}. This is possible due to the parity violation of the weak decay which results in a positron emission preferentially along the $\mu^+$ spin direction at the time of decay. The positron count rate~$N(t)$ is given by:
\begin{equation}
N(t) = N_0 \; {\rm e}^{-t/\tau_{\mu}} \left[1 + A P(t)\right] + N_{Bkg},
\end{equation}
with a mean $\mu^+$ lifetime of ${\tau_{\mu} = 2.197\,\mu{\rm s}}$, where $N_0$ gives the scale of the counted positrons, $N_{\rm Bkg}$ is a time-independent background of uncorrelated events, and $A$ is the observable decay asymmetry.

The experiments were performed in the Meissner state. After zero field cooling to $5$\,K a magnetic field~$B_{\rm ext}=8$\,mT$< \mu_0 H_{{\rm c}_1}$ is applied parallel to the thin-film surfaces and perpendicular to the initial muon spin. In this case the muon-spin polarization $P(t)$ is given by
\begin{equation}
P\left(t\right) = e^{-\sigma^2t^2/2} \int n\left(z\right) \cos \left[\gamma_{\mu} B\left(z\right)t+\phi\right]dz,
\end{equation}
with the initial phase~$\phi$, the muon gyromagnetic ratio~${\gamma_{\mu}=2 \pi \cdot 135.5}$\,MHz/T, and the depolarization rate~$\sigma$ which is a measure of any inhomogeneous local magnetic field distribution at the $\mu^+$ site. The muon stopping distributions $n(z)$ for different energies are simulated using the Monte Carlo code TRIM.SP~\cite{Eckstein91} (see Fig.~\ref{fig:Profile}). The applicability of the TRIM.SP code for low energy muons stopping in matter is demonstrated in Ref.~\onlinecite{Morenzoni01}. Since the thin films have twinned $a$- and $b$-axes, the average magnetic penetration depth~$\lambda_{ab}$ is determined with the present experimental set-up. To analyze the data, the London model profile for thin films was used, resulting from an exponential decay of $B_{\rm ext}$ from both interfaces:
\begin{equation}
B\left(z\right) =
\begin{cases}
B_{\rm ext} \frac{\displaystyle \cosh \left(\frac{z-z_0-d_{\rm sc}/2}{\lambda_{ab}}\right)}{\displaystyle \cosh \left(\frac{d_{\rm sc}/2}{ \lambda_{ab}}\right)} &, z \geq z_0\\
B_{\rm ext} &, z<z_0
\end{cases}
\end{equation}
where $d_{\rm sc}$ is the thickness of the superconducting region, $z$ is the depth along the $c$-axis, and $z_0$ is the effective dead layer at the vacuum interface (see Fig.~\ref{fig:RBS}). This dead layer may be due to the roughness of the surface~\cite{Lindstrom} or to a reduction of the order parameter close to the surface.

\begin{figure}[b]
	\centering
		\includegraphics[width=\columnwidth]{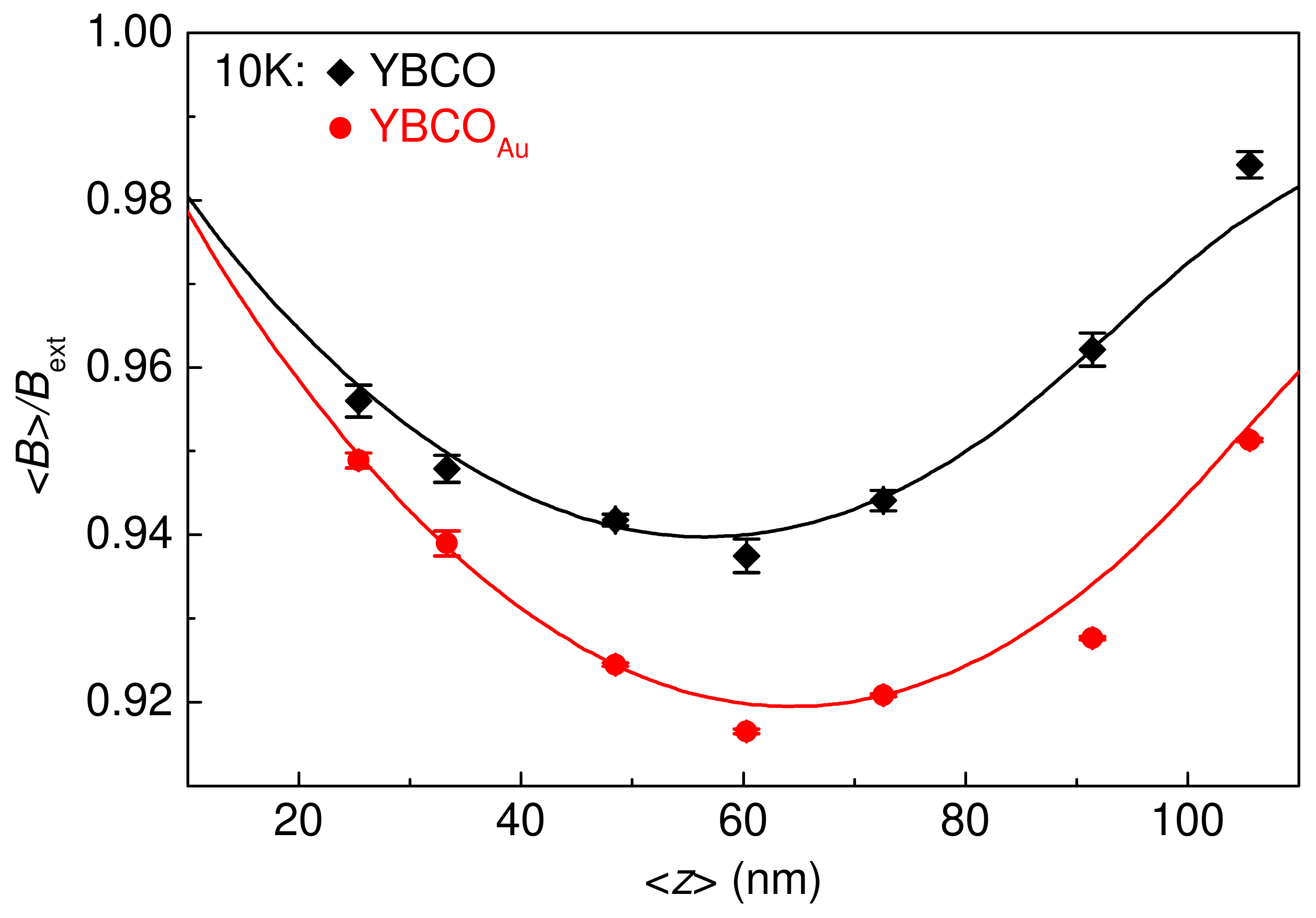}
	\caption{(Color online) The average local magnetic field normalized to the applied field $\langle B \rangle/B_{\rm ext}$ versus the mean muon implantation depth $\langle z \rangle$ measured in an applied magnetic field $B_{\rm ext}=8$\,mT parallel to the film surface at $10$\,K for thin-film YBCO and YBCO$_{\rm Au}$. The solid lines represent the results of the global fits obtained with musrfit~\cite{Suter12}, where seven $\mu$SR spectra with different energies were analyzed simultaneously. The used parameter values are listed in Table~\ref{tab:Tabble1}. The data points are determined from single energy fits using $\langle B \rangle=\int n(z) B(z) {\rm d}z$, where $n(z)$ is the muon implantation profile of the corresponding energy (see Fig.~\ref{fig:Profile}).}
	\label{fig:Penetration}
\end{figure}

By the simultaneous analysis of the $\mu$SR spectra measured at seven different implantation energies at $10$\,K, the Meissner screening profiles of the two sample sets YBCO and YBCO$_{\rm Au}$ were determined (see Fig.~\ref{fig:Penetration}). The resulting values of the parameters are listed in Table~\ref{tab:Tabble1}. The thickness of the superconducting layer~$d_{\rm sc}$ is substantially smaller than the average film thickness~$\overline{d}$ in both sample sets. This confirms the presence of a dead layer at the substrate interface $d_{\rm DL}$ of about ${31}$\,nm for \mbox{YBCO} and about ${18}$\,nm for \mbox{YBCO$_{\rm Au}$}. The interface layer ($d_{\rm DL}$) exhibits only weak superconductivity as mentioned before. No macroscopic Meissner screening is observed in this region. This could be explained by the presence of grain boundaries due to mechanical strain~\cite{Schneidewind95}, leading also to the reduction of $T_{\rm c}$ in thin films compared to single-crystal YBCO. Note that $d_{\rm DL}$ is smaller for the YBCO$_{\rm Au}$ samples, indicating that the Au nanoparticles may relax the mechanical strain. The dead layers to the vacuum interface are surprisingly small in both sets ($z_0 \approx 1$\,nm) compared to single-crystal YBCO ($z_0 \simeq 10$\,nm)~\cite{Hossain12} and other YBCO thin-film measurements ($z_0 \simeq 8$\,nm)~\cite{Jackson00}.

\begin{figure}[t]
	\centering
		\includegraphics[width=1.0\columnwidth]{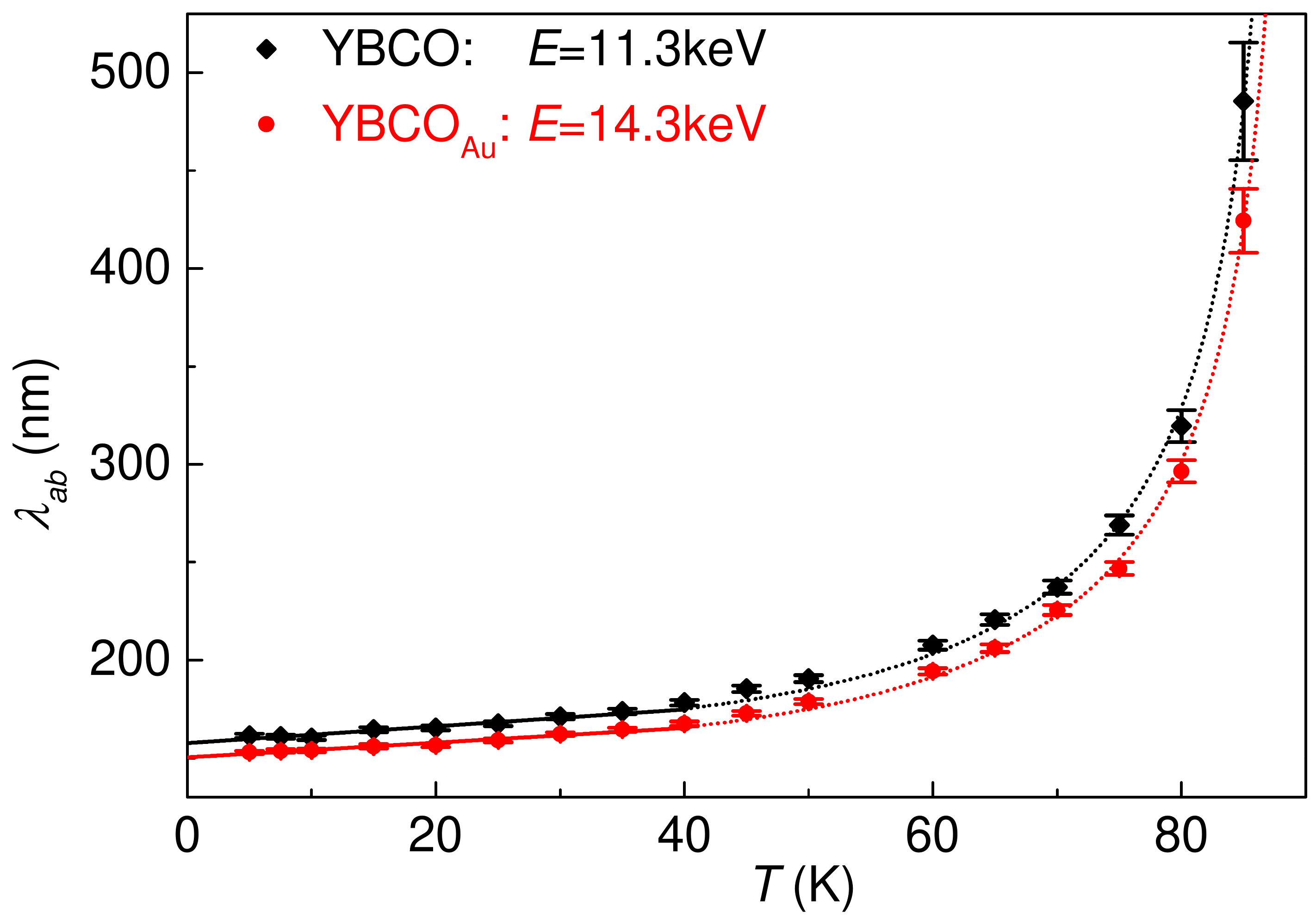}
	\caption{(Color online) The London penetration depth $\lambda_{ab}$ versus temperature $T$ for thin-film YBCO and YBCO$_{\rm Au}$. $\lambda_{ab}$ was determined from ${\langle B \rangle/B_{\rm ext}}$ (Gaussian fit to the $\mu$SR spectra with musrfit~\cite{Suter12}) by using the parameters from the global fit determined at $10$\,K. The applied magnetic field was ${B_{\rm ext}=8}$\,mT. The implantation energies $E$ given in the figure were selected according to the minimum in ${\langle B \rangle/B_{\rm ext}}$. The corresponding solid lines are linear fits to the data in the range ${5-35}$\,K by means of Eq.~(\ref{eq:dwave}). The parameter values are listed in Table~\ref{tab:Tabble1}.  The dotted lines are fits to the power law $\lambda_{ab}(T) = \lambda_{ab}(0)[1-(T/T_{\rm c})^n]^{-1/2}$ with $n\simeq2.7$.}
	\label{fig:Lambda}
\end{figure}

\mbox{YBCO$_{\rm Au}$} screens the magnetic field better, and therefore exhibits a smaller $\lambda_{ab}$ than our pristine YBCO (Fig.~\ref{fig:Penetration}). The temperature dependence of $\lambda_{ab}$ of thin-film YBCO and YBCO$_{\rm Au}$ is presented in Fig.~\ref{fig:Lambda}. These values of $\lambda_{ab}$ were extracted from $\mu$SR spectra measured at the energy corresponding to the minimal average local magnetic field $\langle B \rangle_{\rm min}$ at $10$\,K. Both sample sets show a linear increase in $\lambda_{ab}$ up to $35$\,K which is characteristic for a $d$-wave pairing~\cite{Poole}:
\begin{equation}
\lambda_{ab}(T) = \lambda_{ab}(0{\rm K}) \cdot \left[ 1 + \ln\left(2\right) \cdot k_{\rm B} T / \Delta_0 \right].
\label{eq:dwave}
\end{equation}
The slopes for YBCO and YBCO$_{\rm Au}$ are very similar (${\approx 0.4}$\,nm/K), so the superconducting gaps $\Delta_0$ are the same within experimental uncertainty (see Table~\ref{tab:Tabble1}). The value of $\lambda_{ab}(0{\rm K})$ for \mbox{YBCO$_{\rm Au}$} is smaller compared to \mbox{YBCO}, but still larger than that one of single-crystal YBCO. If the amount of Au nanoparticles is higher (Au seed layer of $3$\,nm instead of $1.8$\,nm) the same effects on $T_{\rm c}$ and $\lambda_{ab}$ where observed. The nominal changes are smaller, since most of the Au nanoparticles remained at the interface to the substrate.

A possible explanation for the reduced magnetic penetration depth in YBCO$_{\rm Au}$ is the diffusion of Au from the nanoparticles into the Cu-O chains (likely during the YBCO deposition), since Au has a relatively high mobility at higher temperatures. A replacement of Cu(1) by Au in the Cu-O chains leads to an increase of the oxygen mobility~\cite{Claus93}. An enhanced oxygen mobility favors longer Cu-O chain segments leading to in increase of the charge carrier transfer from the Cu-O chains to the CuO$_2$ planes~\cite{Aligia94}. Since more charge carriers can contribute to superconductivity, $n_{\rm s}$ increases while $\lambda_{ab}$ decreases. In this case the $c$-axis lattice constant should be slightly increased~\cite{Cieplak1990}. This, however, was not observed in the investigated samples.

Another possibility is that the presence of Au nanoparticles in the superconducting region of YBCO lowers the defect density and leads therefore to the observed reduction of $\lambda_{ab}$ in YBCO$_{\rm Au}$. A reduced defect density could originate from a condensation of defects at the Au nanoparticles, so that the CuO$_2$ planes and the structure are less disturbed. To  evaluate the defect density we take the $c$-axis lattice constant as an indication. In YBCO$_{\rm Au}$, we observe a slightly smaller $c$-axis lattice constant ($c_{\rm YBCO_{Au}}=1.1688(2)$\,nm $< c_{\rm YBCO}=1.1699(2)$\,nm). YBCO single crystals have an even smaller $c$-axis lattice constant ($c_{\rm bulk}=1.167(1)$\,nm)~\cite{Benzi04}. A lower defect density is generally expected in single-crystal YBCO compared to thin-film YBCO which is consistent with the trend of the $c$-axis lattice constants. From this point of view YBCO$_{\rm Au}$ has a lower defect density than our pristine YBCO, but a higher one compared to single-crystal YBCO. Since single-crystal YBCO has also a smaller $\lambda_{ab}$ and a higher $T_{\rm c}$ compared to our YBCO and YBCO$_{\rm Au}$ samples, a reduced defect density is a possible reason for the slightly different values of $T_{\rm c}$ and $\lambda_{ab}(0)$ in YBCO$_{\rm Au}$ compared to pure YBCO films.

In summary, we have shown by using LE-$\mu$SR that the increase of $T_{\rm c}$ in thin-film YBCO containing Au nanoparticles is accompanied by a reduction of the average magnetic penetration depth $\lambda_{ab}$. We attribute this reduction to a lowered defect density. The diffusion of Au into the Cu-O chains may also contribute to the observed changes of $T_{\rm c}$ and $n_{\rm s}$. In previous studies on YBCO, Cu was exchanged by Au within the Cu-O chains to obtain an enhanced $T_{\rm c}$. Our results indicate that the presence of Au nanoparticles has a similar effect in thin-film YBCO.

We acknowledge Hans-Peter Weber for his excellent technical support. This work was partly supported by the Swiss National Science Foundation and the Landesgraduiertenf\"oderung Th\"uringen.


\begin{thebibliography}{00}
\bibitem{Alloul09}
H.\ Alloul, J.\ Bobroff, M.\ Gabay, and P.\ J.\ Hirschfeld, Rev.\ Mod.\ Phys. \textbf{81}, 45 (2009).
\bibitem{Aligia94}
A.\ A.\ Aligia and J.\ Garces, Phys. Rev. B \textbf{49}, 524 (1994).
\bibitem{Shlyk02}
L.\ Shlyk, G.\ Krabbes, G.\ Fuchs, G.\ St\"over, S.\ Gruss, and K.\ Nenkov, Physica\ C \textbf{377}, 437 (2002).
\bibitem{Cieplak1990}
M.\ Z.\ Cieplak, G.\ Xiao, C.\ L.\ Chien, J.\ K.\ Stalick, and J.\ J.\ Rhyne, Appl.\ Phys.\ Lett. \textbf{57}, 934 (1990).
\bibitem{Cieplak1990b}
M.\ Z.\ Cieplak, G.\ Xiao, C.\ L.\ Chien, A.\ Bakhshai, D.\ Artymowicz, W.\ Bryden, J.\ K.\ Stalick, and J.\ J.\ Rhyne, Phys.\ Rev.\ B \textbf{42}, 6200 (1990).
\bibitem{Welp93}
U.\ Welp, S.\ Fleshler, W.\ K.\  Kwok, J.\ Downey, G.\ W.\ Crabtree, H.\ Claus,A.\ Erb, and G.\ M\"uller-Vogt, Phys.\ Rev.\ B \textbf{47}, 12369 (1993).
\bibitem{Michalowski12}
C.\ Katzer, M.\ Westerhausen, I.\ Uschmann, F.\ Schmidl, U.\ H\"ubner, and P.\ Seidel, Supercond.\ Sci.\ Technol. \textbf{26}, 125008 (2013).
\bibitem{Katzer11}
C.\ Katzer, M.\ Schmidt, P.\ Michalowski, D.\ Kuhwald, F.\ Schmidl, V.\ Grosse, S.\ Treiber, C.\ Stahl, J.\ Albrecht, U.\ H\"ubner, A.\ Undisz, M.\ Rettenmayr, G.\ Sch\"utz, and P.\ Seidel, EPL \textbf{95}, 68005 (2011).
\bibitem{Katzer12}
C.\ Katzer, C.\ Stahl, P.\ Michalowski, S.\ Treiber, F.\ Schmidl, P.\ Seidel, J.\ Albrecht, and G.\ Sch\"utz, New\ J.\ Phys. \textbf{15}, 113029 (2013).
\bibitem{Doebeli98}
M.\ D\"obeli, J.\ Phys.: Condens.\ Matter \textbf{20}, 264010 (2008). 
\bibitem{Doolittle86}
L.\ R.\ Doolittle, Nucl.\ Instrum.\ Methods Phys.\ Res.\ B \textbf{15}, 227 (1986).
\bibitem{Schneidewind95}
H.\ Schneidewind, F.\ Schmidl, S.\ Linzen, and P.\ Seidel, Physica\ C \textbf{250}, 191 (1995).
\bibitem{Morenzoni04}
E.\ Morenzoni, T.\ Prokscha, A.\ Suter, H.\ Luetkens, and R.\ Khasanov , J.\ Phys.: Cond.\ Matt. \textbf{16}, S4583 (2004).
\bibitem{Prokscha08}
T.\ Prokscha, E.\ Morenzoni, K.\ Deiters, F.\ Foroughi, D.\ George, R.\ Kobler, A.\ Suter, and V.\ Vrankovic, Nucl.\ Instrum.\ Methods Phys.\ Res., Sect.\ A \textbf{595}, 317 (2008).
\bibitem{Eckstein91}
W.\ Eckstein, \textit{Computer Simulation of Ion-Solid Interactions} (Springer, Berlin, 1991).
\bibitem{Yaouanc}
A.\ Yaouanc and P.\ Dalmas de Reotier, \textit{Muon Spin Rotation, Relaxation, and Resonance: Applications to Condensed Matter} (Oxford University Press, Oxford, 2011); V.\ P.\ Smilga and Yu.\ M.\ Belousov, \textit{The Muon Method in Science} (Nova Science Publishers, New York, 1994).
\bibitem{Morenzoni01}
E.\ Morenzoni, H.\ Gl\"uckler, T.\ Prokscha, R.\ Khasanov, H.\ Luetkens, M.\ Birke, E.\ M.\ Forgan, Ch.\ Niedermayer, and M.\ Pleines, Nucl.\ Instr.\ Meth.\ B \textbf{192}, 254 (2002).
\bibitem{Lindstrom}
M.\ Lindstrom, B.\ Wetton, and R.\ Kiefl, J.\ Eng.\ Math., DOI 10.1007/s10665-013-9640-y.
\bibitem{Suter12}
A.\ Suter and B.\ M.\ Wojek, Physics Procedia \textbf{30}, 69 (2012).
\bibitem{Hossain12}
R.\ F.\ Kiefl, M.\ D.\ Hossain, B.\ M.\ Wojek, S.\ R.\ Dunsiger, G.\ D.\ Morris, T.\ Prokscha, Z.\ Salman, J.\ Baglo, D.\ A.\ Bonn, R.\ Liang, W.\ N.\ Hardy, A.\ Suter, and E.\ Morenzoni, Phys.\ Rev.\ B \textbf{81}, 180502(R) (2010).
\bibitem{Jackson00}
T.\ J.\ Jackson, T.\ M.\ Riseman, E.\ M.\ Forgan, H.\ Gl\"uckler, T.\ Prokscha, E.\ Morenzoni, M.\ Pleines, Ch.\ Niedermayer, G.\ Schatz, H.\ Luetkens, and J.\ Litterst, Phys.\ Rev.\ Lett. \textbf{84}, 4958 (2000).
\bibitem{Poole}
C.\ P.\ Poole Jr., H.\ A.\ Farach, R.\ J.\ Creswick, and R.\ Prozorov, \textit{Superconductivity} (Academic Press, London, 2007)
\bibitem{Claus93}
H.\ Claus, S.\ Yang, H.\ K.\ Viswanathan, G.\ W.\ Crabtree, J.\ W.\ Downey, and B.\ W.\ Veal, Physica\ C \textbf{213}, 185 (1993).
\bibitem{Benzi04}
P.\ Benzi, E.\ Bottizzo, and N.\ Rizzi, J.\ Cryst.\ Growth \textbf{269}, 625 (2004).
\end{thebibliography}
\end{document}